# Optimized Operation of Available Energy Resources Based on Energy Consumption

Parvathy S
*Robert Bosch*
Bangalore,India
parvathy2202sobha@gmail.com

Nita R Patne
*Electrical Engineering Department*
VNIT, Nagpur, India
nrpatne.eee@vnit.ac.in

*Abstract*— Energy consumption and energy ananlytics has gained increased focus and consideration in industrial applications especially process lines to upgrade their performance and efficiency in the competitive world. A competent analytics method will be highly advantageous to provide the correct direction of energy saving for an industry. Energy analytics method was introduced into energy consumption time analysis model and is developed in this paper. Energy consumption of a camplate production plant was analysed as a case study. The result shows that energy utilization is dependent on the time of operation of the equipments in the plant. Energy sparing obtained via technical innovation in unit process maybe misplaced due to the expanding time of operation within the plant. Energy loss and the distribution of energy loss in the camplate production plant were analysed and the probable energy saving methods were identified from the results.

## I. INTRODUCTION

The format of power generation capacity is planned to capture the peak demand to guarantee unwavering quality in supply. India still faces challenges in assembling its developing request for control and solid supply of growing energy demand, the World Bank said in a report. With a growing population, quick urbanization and an economy that's anticipated to develop at a normal rate of 7% each year, request for power in India will nearly triple between 2018 & 2040, World Bank said, citing projections from International Energy Agency [1]-[3].

In this scenario, this drastic increase in the energy consumption of the industrial sector in the recent years possess a serious threat to the power grid. With the same production rate or slight increase in production, the energy demand has risen in this sector. The production rate cannot be manipulated to address this issue. On the other hand, measures must be made in the efficient handling of the production lines in order to achieve energy efficiency in industries [3]. Variables influencing energy utilization of an industry must be evaluated through their life cycle. Those factors majorly include the equipments employed in production line and their time of operation. These components have a coordinate effect on the energy effectiveness of buildings and comes about in an expanded outflows of carbon dioxide[4].

Energy optimization in the production line can be achieved through various ways. Many parameters/factors decide the energy consumption hence the optimization of energy is a tedious process Energy optimization of industrial sector has been studied since 1950s. Conventional analysis employees Life Cycle Costing (LCC) and Life Cycle Assessment (LCA). These have been employed to assess the environmental impacts and costs of the energy of industrial process lines since a very long time [5-7].

Barry Hyman and Tracy Reed has developed a more generic approach to assess the energy-intensity of manufacture process lines [9]. Many other methods including Standard Material Flow Diagram was developed to analyze the energy usage and production in various industrial fields [10-11]. But all these conventional methods cannot be implemented in all developing countries. In these nations one major aspect to be considered is the constrained information of the workforce within the generation unit. Significant extent of work drive driving or regulating the loads in units is semi-skilled, this may quicken toward energy wastage through incorrect operation of equipments and device breakdown. Subsequently there are tall chances of energy wastage within the generation unit.

The work presented in this paper introduces a competent analytics method including the operating time of equipments in the process line. This strategy will be exceedingly profitable to supply the right course of energy sparing for an industry by lessening energy wastage. The paper is organized as follows: Section II defines the objective of the work. Section III details the methodology followed in the work. Section IV includes the discussions and recommendations from analysis. Section V corporates the conclusions drawn from the analysis.

## II. OBJECTIVE

The existing literature[12]-[15] commonly does not include the consumption time owing to the lack of skilful workers in the production line. Authors have employed various optimization models and different learning algorithms in production line [16-22]. But the use of various statistical analysis in are not explore in depth. The work presented in this paper compares and analyze the ideal operating time to the actual operating time of various equipments in the production line. Within the work, a diesel infusion pump fabricating unit in India is the test location considered. The operation of Camplate Process Line in the plant is investigated. The collected data from various installed meters are analyzed to find anomalous operating points in the process line.

## III. SYSTEM DETAILS

Camplate Process Line

A cam is a turning or drifting piece in a mechanical linkage utilized particularly in changing rotating movement into straight movement. It is frequently a portion of a pivoting wheel (e.g. an eccentric wheel) or shaft (e.g. a barrel with an sporadic shape) that beats a lever at one or more focuses on its circular way. In camplate process, raw camplate from foundry are reformed to finished camplate through eight distinct processes as shown in Fig. 1. This figure describes the process line, all the components in the process line, the function of each components, the ideal operating time of equipments and their rating. The major components include:

1. Kadia – employed for honing the raw camplate. It is an acerbic machining process by which precision surface is developed on the raw camplate by rubbing9. an casuistic material against it along a controlled path
2. Jumat – employed for grinding the raw camplate
3. ECM – employed for grinding off the claw base of the raw camplate
4. MS-19 – employed for grinding the can profile
5. Supfina – employed for shim face grinding
6. Cleaning – cleans the finished camplate
7. Shot peening – employed for relieving stress developed while grinding the camplate
8. Trowalising – employed for deburring the sharp edges of the camplate

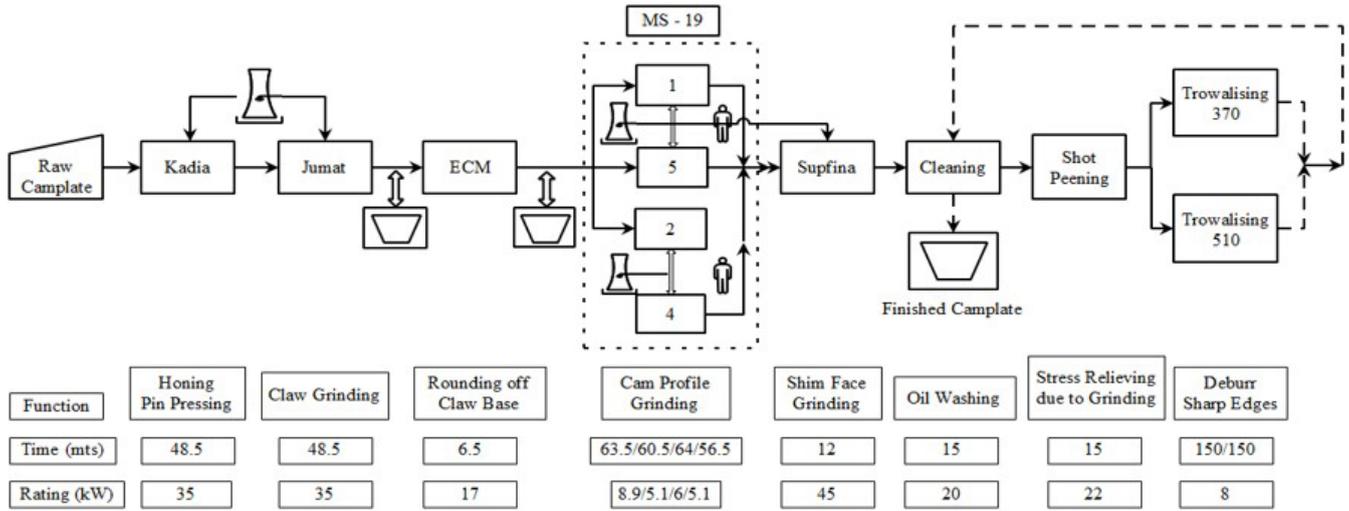

Fig. 1. Detailing of camplate process line

From eight processes, two processes with the least operating time and two processes with the highest operating time are further investigated which includes shot peening, cascade camplate cleaning, trowalising (two loads: trowalising 510 and trowalising 370).

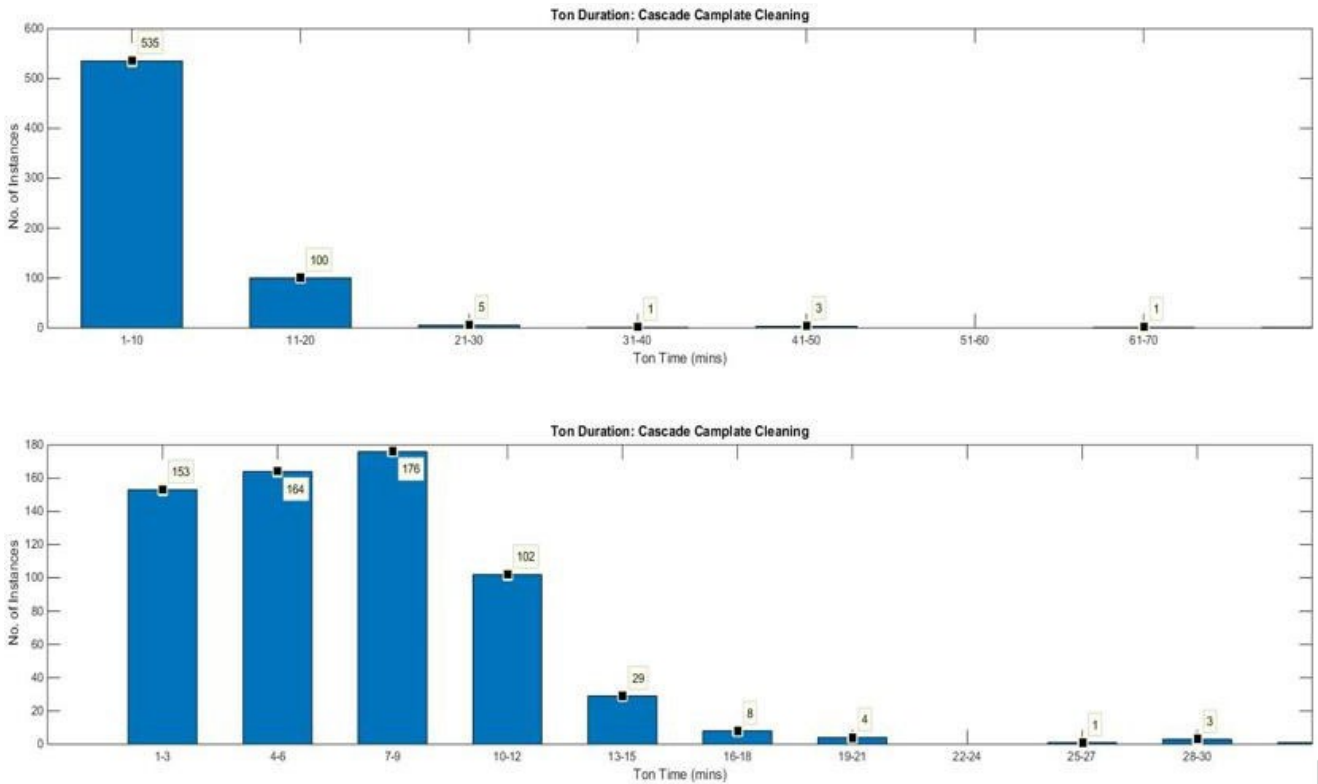

Fig. 2. $T_{on}$ instances of cascade camplate cleaning.

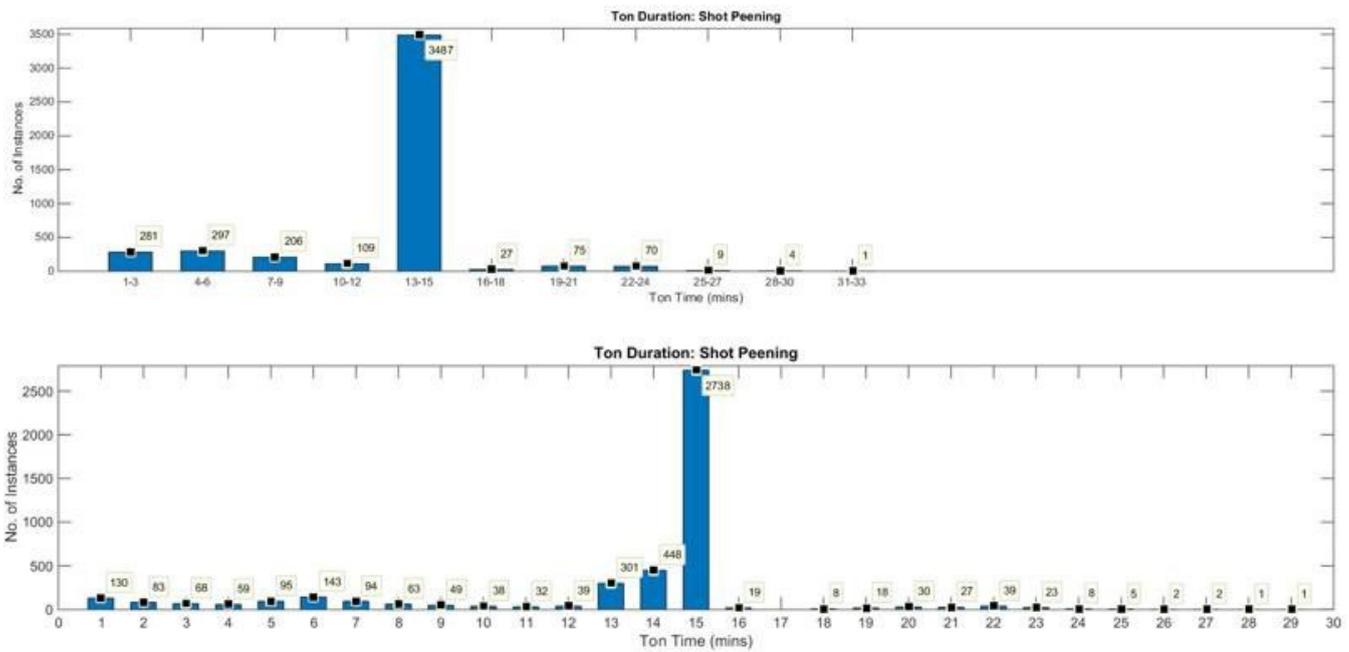

Fig. 3. $T_{on}$ instances of shot peening.

## IV. METHODOLOGY

Camplate process is a batch process. hence every process in the line should adhere to the rated process time. To investigate this, the $T_{on}$ time and number of instances corresponding to this $T_{on}$ time are determined and analyzed. The results of Statistical analysis of data are explained in the below sections.

For camplate cleaning process, the ideal $T_{on}$ time is 15 minutes. Fig. 2 illustrates the results of $T_{on}$ analysis. It's observed that the number of instances with $T_{on}$ duration comparable to ideal $T_{on}$ time is less i.e. 37 instances (out of 645 instances) with operating time between 13-18 minutes. In majority cases the $T_{on}$ time is much less than the ideal operating time. Hence occurrences of operational anomalies are detected in the process load.

For shot peening, the ideal $T_{on}$ time is 15 minutes. Fig. 3 shows the results of $T_{on}$ analysis. It's been observed that majority of instances (2738 out of 4589 instances) complete the operation with in rated operating time. Hence currently, no operational anomalies are present in the process load.

For trowalising 370, the ideal $T_{on}$ time is 150 minutes. Fig.4 illustrates the results of $T_{on}$ analysis. It's observed that the number of instances with $T_{on}$ duration comparable to ideal $T_{on}$ time is much less i.e. 5 instances (out of 1526 instances) with operating time between 151-155 minutes. In majority cases the $T_{on}$ time is much less than the ideal operating time. Hence occurrences of operational anomalies are detected in the process load. For trowalising 510, the ideal $T_{on}$ time is 150 minutes.

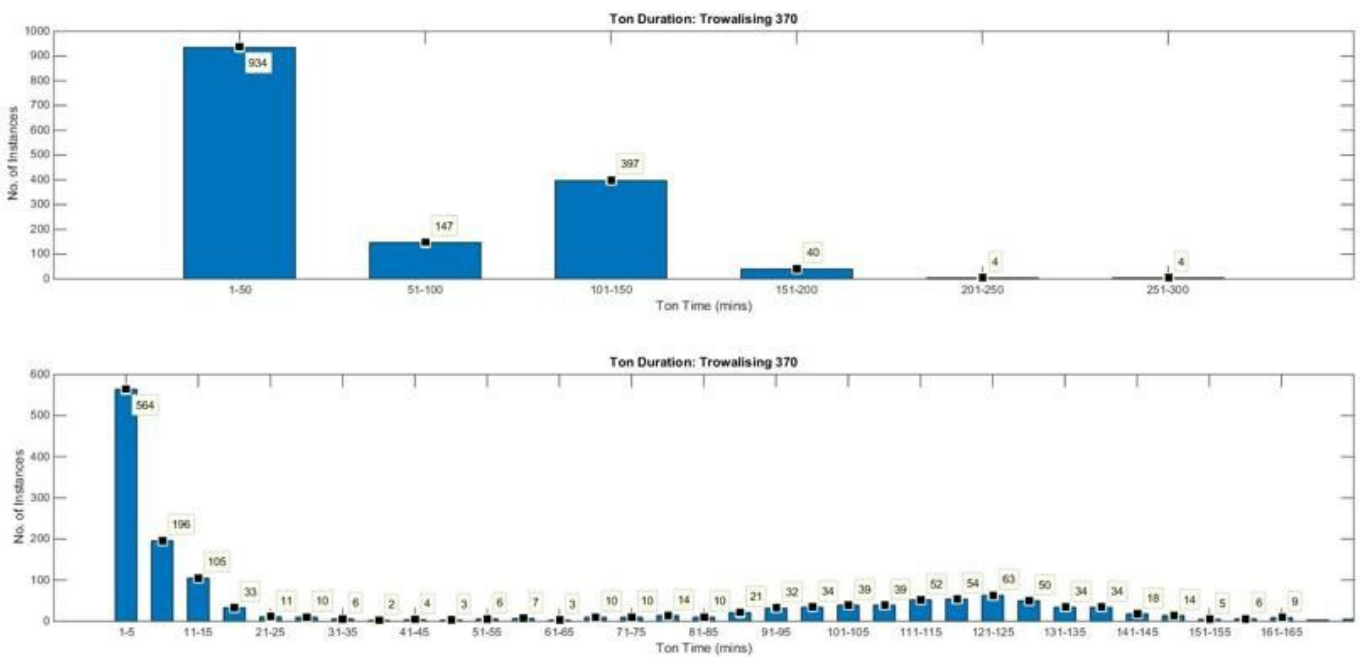

Fig. 4. $T_{on}$ instances of trowalising 370.

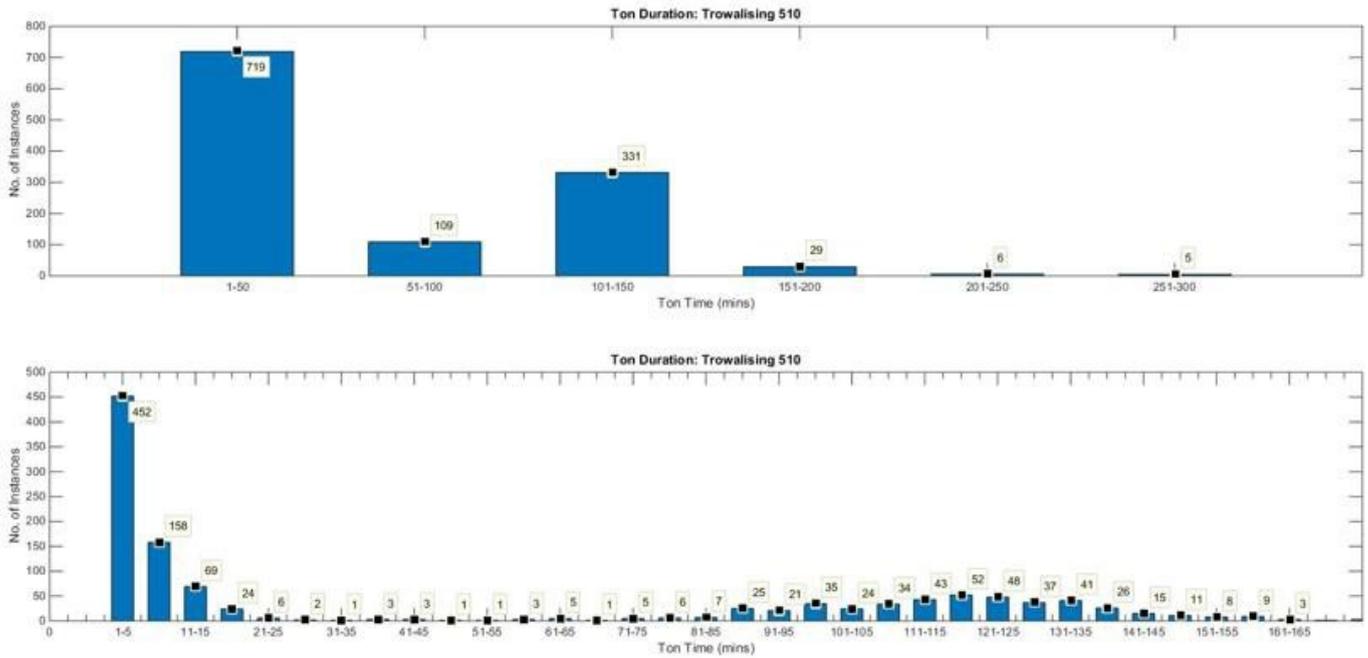

Fig. 5. $T_{on}$ instances of trowalising 510.

Fig. 5 illustrates the results of $T_{on}$ analysis. It's observed that the number of instances with $T_{on}$ duration comparable to ideal $T_{on}$ time is much less i.e. 8 instances (out of 1199 instances) with operating time between 151-155 minutes. In majority cases the $T_{on}$ time is much less than the ideal operating time. Hence occurrences of operational anomalies are detected in the process load. Apart from shot peening, the three other loads exhibit abnormalities in operating duration.

For further analysis the scatter plots of the operating time of the loads are analyzed. In the scatter plots illustrated,
$\sigma_E$ represents mean of load energy consumption
$\mu_t$ represents mean of $T_{on}$ time
$\sigma_t$ represents standard deviation of $T_{on}$ time
$\mu_E$ represents mean of load energy consumption

Fig. 6 illustrates the scatter plot of cascade camplate cleaning. The presence of anomaly is validated by the mean and standard deviation of process $T_{on}$ duration. The y axis represents the energy consumption of the load during corresponding $T_{on}$ time. Undoubtedly, it's the energy wasted during prolonged operating time of the load.

The average energy consumption is 2.43 kWh where the maximum consumption goes till 7.5 kWh with a considerable standard deviation in energy consumption (of 2.89 kWh). Fig. 7 illustrates the scatter plot of shot peening. The standard deviation of energy consumption is as low as 0.87 and the mean of operating time is 13 minutes, close to ideal operating duration of 15 minutes. Hence shoot peening pursues normal operating duration.

The scatter plot of trowalising 370 is illustrated by Fig. 8. The mean and standard deviation of $T_{on}$ duration confirms abnormal operation in the system. The mean of $T_{on}$ duration is 50 minutes where the rated $T_{on}$ time is 150 minutes. The process shows abnormal operating pattern. The lowered operating duration leads to reduced production quality. Prior to energy wastage production quality is affected by this anomaly. Fig. 9 illustrates the scatter plot of trowalising 510. The mean of $T_{on}$ time is 130 minutes where ideal duration is 150 minutes. Operational anomalies are shown by the load like trowalising 370.

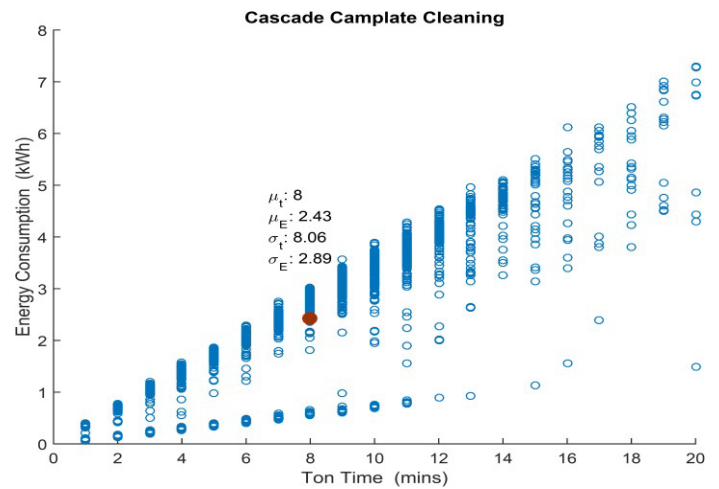

Fig. 6. Scatter Plot of cascade camplate cleaning.

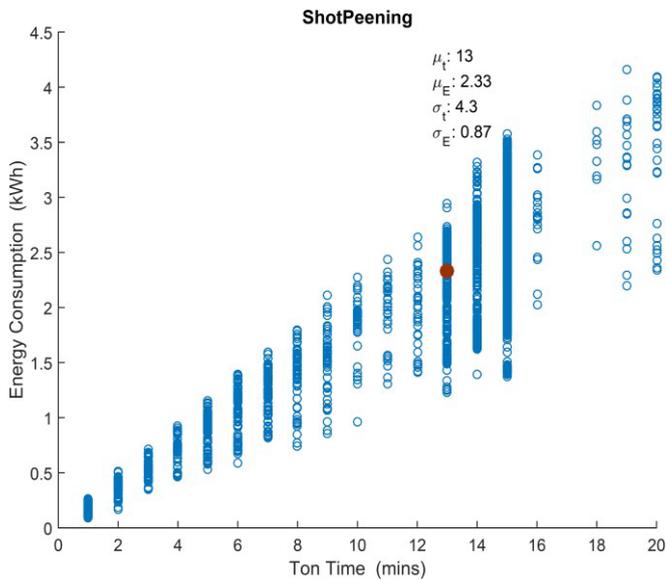
Fig. 7. Scatter Plot of shot peening.

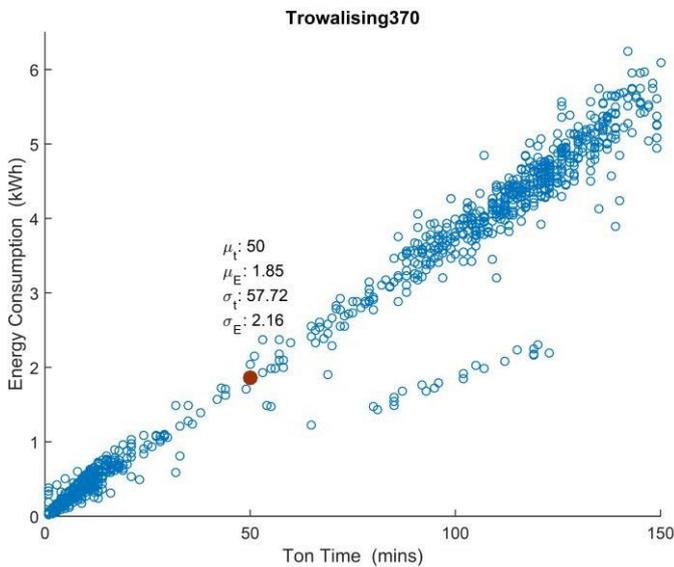
Fig. 8. Scatter Plot of trowalising 370.

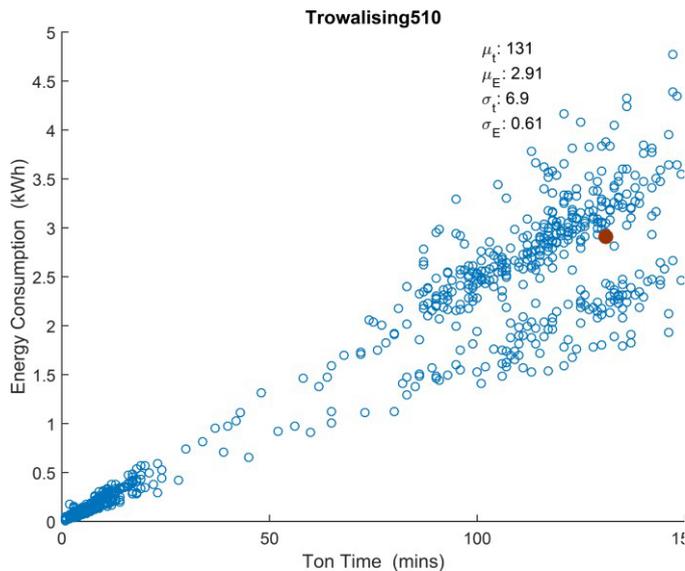
Fig. 9. Scatter Plot of trowalising 510.

## V. DISCUSSION AND RECOMMENDATION

Anomalous operations are detected in three process loads. While reporting to the FM, the probable source of abnormalities suggested includes,

- the process being operated in under load or over load
- enhanced machine down time during process cycle
- human error during manual operations of the load
- machine exceeding thermal limits
- supply problems
- sudden change in batch size

Proposed remedies include,

- routine maintenance checks
- Keep suggested batch size
- aversion of sudden changes in batch size (in case to meet production target)
- standardization of work: fool proofing, bench marking

## VI. CONCLUSION

The camplate process line comprises of several stages. The erroneous operating points at different stages are determined through data analysis. These findings are reported to the Facility Manager of the plant. The probable causes of these abnormalities and the remedies for these faulty operations are communicated. A little increase in operational proficiency comes about in noteworthy energy savings. Thus the proposed strategy progresses the generation quality, amount with diminished operational cost.